\documentclass[12pt]{article}
\usepackage{axodraw}

\catcode`@=12
\begin{document}
\thispagestyle{empty}
\begin{flushright} 
UCRHEP-T371\\ 
March 2004\
\end{flushright}
\vspace{0.5in}
\begin{center}
{\LARGE	\bf Exotic Fermions and Bosons\\ in the Quartification Model\\}
\vspace{1.5in}
{\bf Shao-Long Chen and Ernest Ma\\}
\vspace{0.2in}
{\sl Physics Department, University of California, Riverside, 
California 92521\\}
\vspace{1.5in}
\end{center}
\begin{abstract}\
Exotic fermions of half-integral charges at the TeV energy scale are 
predicted by the quartification model of Babu, Ma, and Willenbrock. 
We add to these one copy of their scalar analogs and discuss the ensuing 
phenomenological implications, i.e. radiative contributions to lepton 
masses and flavor-changing leptonic decays.
\end{abstract}

\newpage
\baselineskip 24pt

In the recently proposed \cite{bmw} model of $SU(3)^4$ quartification, 
the low-energy gauge group of particle interactions is $SU(3)_C \times 
SU(2)_L \times U(1)_Y \times SU(2)_l$ and the particle content is that 
of the nonsupersymmetric standard model with two Higgs doublets, and 
three families of exotic fermions.  Whereas the former are singlets under 
the new gauge symmetry $SU(2)_l$, the latter are doublets, i.e.
\begin{equation}
\pmatrix {x \cr y} \sim (1,2,0;2), ~~~ x^c \sim (1,1,-{1 \over 2}; 2), 
~~~ y^c \sim (1,1,{1 \over 2}; 2).
\end{equation}
Because they have $\pm 1/2$ electric charges, they have been named 
``hemions''.

The $SU(2)_l$ gauge symmetry is the unbroken remnant of a leptonic color 
$SU(3)_l$ \cite{fl90} which combines with the familiar $SU(3)_C \times SU(3)_L 
\times SU(3)_R$ to form $SU(3)^4$ at the quartification scale \cite{bmw}. 
The exotic hemions are required to be at the TeV energy scale, or else the 
gauge couplings would not unify.  On the other hand, they do not interact 
directly with quarks or leptons, so they are produced only through the 
electroweak gauge bosons, i.e. $W^\pm$, $Z$, and $\gamma$.  Since the 
hemions are $SU(2)_l$ doublets, they couple only in pairs, and because 
they are also fermions, they have Yukawa couplings only to the Higgs 
scalars.  Suppose we now add one set of scalar hemions $(\tilde x, \tilde y)$, 
$\tilde x^c$, and $\tilde y^c$ transforming in the same way as $(x,y)$, $x^c$, 
and $y^c$ of Eq.~(1), then Yukawa couplings such as $(xe-y\nu)\tilde y^c$, 
$(\tilde x e - \tilde y \nu)y^c$, and $x^c \tilde x^c e^c$  become possible. 
They would induce radiative contributions to lepton masses as well as 
flavor-changing leptonic decays to one-loop order, which may be observed 
experimentally.

The evolution of the gauge couplings of $SU(3)_C$, $SU(2)_L$, and $U(1)_Y$ 
are given in one-loop order by the renormalization-group equation
\begin{equation}
{1 \over \alpha_i(M_1)} - {1 \over \alpha_i(M_2)} = {b_i \over 2 \pi} 
\ln {M_2 \over M_1},
\end{equation}
where
\begin{eqnarray}
SU(3)_C &:& b_s = -11 + {4 \over 3} N_f = -7, \\ 
SU(2)_L &:& b_2 = -{22 \over 3} + 2 N_f + {1 \over 6} N_H = -1, \\ 
U(1)_Y &:& b_Y = {26 \over 9} N_f + {1 \over 6} N_H = 9,
\end{eqnarray}
between the unification scale $M_U$ and the scale $M_X$ at which the hemions 
become massive, and
\begin{eqnarray}
SU(3)_C &:& b_s = -11 + {4 \over 3} N_f = -7, \\ 
SU(2)_L &:& b_2 = -{22 \over 3} + {4 \over 3} N_f + {1 \over 6} N_H = -3, \\ 
U(1)_Y &:& b_Y = {20 \over 9} N_f + {1 \over 6} N_H = 7,
\end{eqnarray}
between $M_X$ and $M_Z$.  The number of families $N_f$ is three and the 
number of Higgs doublets $N_H$ is assumed to be two.  Recognizing that 
$\sin^2 \theta_W = 1/3$ at $M_U$ so that $\alpha_Y$ is normalized to 
$\alpha_U/2$ at $M_U$, we obtain
\begin{eqnarray}
\ln {M_X \over M_Z} &=& {\pi \over 17} \left( {6 \over \alpha_Y} - 
{23 \over \alpha_2} + {11 \over \alpha_s} \right), \\ 
\ln {M_U \over M_Z} &=& {2 \pi \over 17} \left( {1 \over \alpha_Y} - 
{1 \over \alpha_2} - {1 \over \alpha_s} \right).
\end{eqnarray}
Using the input \cite {pdg}
\begin{eqnarray}
\alpha_2 (M_Z) &=& (\sqrt 2/\pi) G_F M_W^2 = 0.0340, \\ 
\alpha_Y (M_Z) &=& \alpha_2 (M_Z) \tan^2 \theta_W = 0.0102, \\ 
\alpha_s (M_Z) &=& 0.1172,
\end{eqnarray}
we then have
\begin{equation}
{M_X \over M_Z} = 2.8, ~~~ {M_U \over M_Z} = 4.4 \times 10^9,
\end{equation}
which is the case of Ref.~[1].

Instead of $N_H = 2$ assumed in Ref.~[1], let us take $N_H = 1$, 
but also add one set of the scalar hemions.  Equations (9) and (10) are 
then changed to
\begin{eqnarray}
\ln {M_X \over M_Z} &=& {\pi \over 119} \left( {37 \over \alpha_Y} - 
{139 \over \alpha_2} + {65 \over \alpha_s} \right), \\ 
\ln {M_U \over M_Z} &=& {6 \pi \over 629} \left( {37 \over \alpha_Y} - 
{105 \over \alpha_2} + {31 \over \alpha_s} \right),
\end{eqnarray}
yielding instead
\begin{equation}
{M_X \over M_Z} = 11.9, ~~~ {M_U \over M_Z} = 2.9 \times 10^{10}.
\end{equation}
Thus $M_X$ and $M_U$ have been shifted upward, but the former is still of 
order 1 TeV and should be accessible to future experimental verification.

In the presence of both hemions and their scalar counterparts (shemions), 
the following Yukawa couplings are allowed under $SU(3)_C \times SU(2)_L 
\times U(1)_Y \times SU(2)_l$:
\begin{eqnarray}
{\cal L}_Y &=& f_1 (\nu y - e x) \tilde y^c + f_2 (\nu \tilde y - e \tilde x) 
y^c + f_3 e^c x^c \tilde x^c \nonumber \\ 
&+& f_4 \nu^c (x \tilde y - y \tilde x) + f_5 \nu^c x^c \tilde y^c + 
f_6 \nu^c y^c \tilde x^c + H.c.,
\end{eqnarray}
where each $SU(2)_l$ invariant of the form $x \tilde y^c$ means 
$x_1 \tilde y_2^c - x_2 \tilde y_1^c$.  Radiative contributions to 
various lepton masses become possible, as shown below.

\noindent (1) Charged leptons:

\begin{figure}[htb]
\begin{center}
\begin{picture}(360,120)(0,0)
\ArrowLine(20,10)(100,10)
\ArrowLine(180,10)(100,10)
\ArrowLine(180,10)(260,10)
\ArrowLine(340,10)(260,10)
\DashArrowArc(180,10)(80,90,180){4}
\DashArrowArcn(180,10)(80,90,0){4}
\Text(60,0)[]{$e$}
\Text(300,0)[]{$e^c$}
\Text(140,0)[]{$x,y^c$}
\Text(220,0)[]{$x^c$}
\Text(180,10)[]{$\times$}
\Text(180,90)[]{$\times$}
\Text(105,70)[]{$\tilde y^c,\tilde x$}
\Text(250,70)[]{$\tilde x^c$}

\end{picture}
\end{center}
\caption{One-loop contributions to the charged-lepton mass matrix.}
\end{figure}
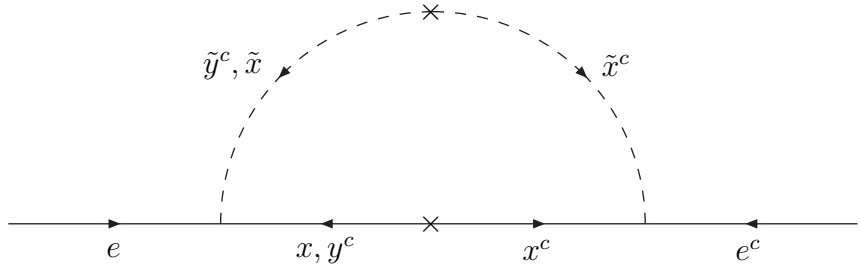

\noindent These contributions are of order
\begin{equation}
m_e = {f_{1,2} f_3 \over 16 \pi^2} m_{eff},
\end{equation}
where $m_{eff}$ is a function of the hemion and shemion masses and their 
couplings.

\noindent (2) Neutrinos:  Contributions to both Dirac and Majorana masses 
are present.  The former comes from the analog of Fig.~1 with couplings 
$f_1 f_{4,6}$ and $f_2 f_{4,5}$ and the latter in the case of $\nu$ is 
depicted below.
 
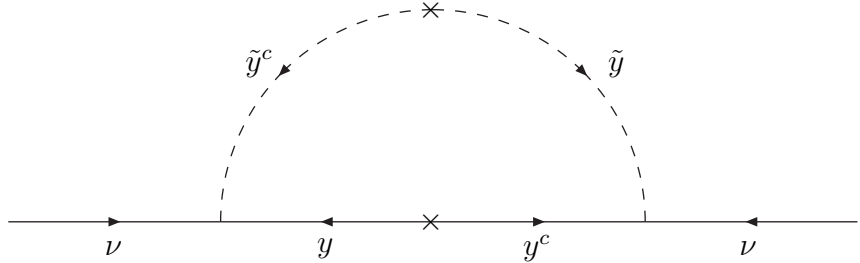
\begin{figure}[htb]
\begin{center}
\begin{picture}(360,120)(0,0)
\ArrowLine(20,10)(100,10)
\ArrowLine(180,10)(100,10)
\ArrowLine(180,10)(260,10)
\ArrowLine(340,10)(260,10)
\DashArrowArc(180,10)(80,90,180){4}
\DashArrowArcn(180,10)(80,90,0){4}
\Text(60,0)[]{$\nu$}
\Text(300,0)[]{$\nu$}
\Text(140,0)[]{$y$}
\Text(220,0)[]{$y^c$}
\Text(180,10)[]{$\times$}
\Text(180,90)[]{$\times$}
\Text(115,70)[]{$\tilde y^c$}
\Text(250,70)[]{$\tilde y$}

\end{picture}
\end{center}
\caption{One-loop contributions to the Majorana neutrino mass matrix.}
\end{figure}

\noindent This contribution is of order
\begin{equation}
m_\nu = {f_1 f_2 \over 16 \pi^2} m'_{eff},
\end{equation}
where $m'_{eff}$ is typically smaller than $m_{eff}$ of Eq.~(19) by at least 
an order of magnitude.  Similar contributions exist in the case of $\nu^c$.

These radiative corrections are not directly observable, because they simply 
add to the already existing tree-level charged-lepton and neutrino masses 
\cite{bmw}.  On the other hand, the same interactions also induce 
flavor-changing leptonic decays.  A typical diagram is shown below.

\begin{figure}[htb]
\begin{center}
\begin{picture}(360,160)(0,0)
\ArrowLine(20,50)(100,50)
\ArrowLine(180,50)(100,50)
\ArrowLine(260,50)(180,50)
\ArrowLine(260,50)(340,50)
\DashArrowArc(180,50)(80,0,180){4}
\Photon(180,50)(180,2){4}{4}
\Text(60,40)[]{$e^c$}
\Text(300,40)[]{$e^c$}
\Text(140,40)[]{$x^c$}
\Text(220,40)[]{$x^c$}
\Text(180,140)[]{$\tilde x^c$}
\Text(180,-7)[]{$\gamma$}

\end{picture}
\end{center}
\caption{Typical diagram for flavor-changing leptonic decay.}
\end{figure}
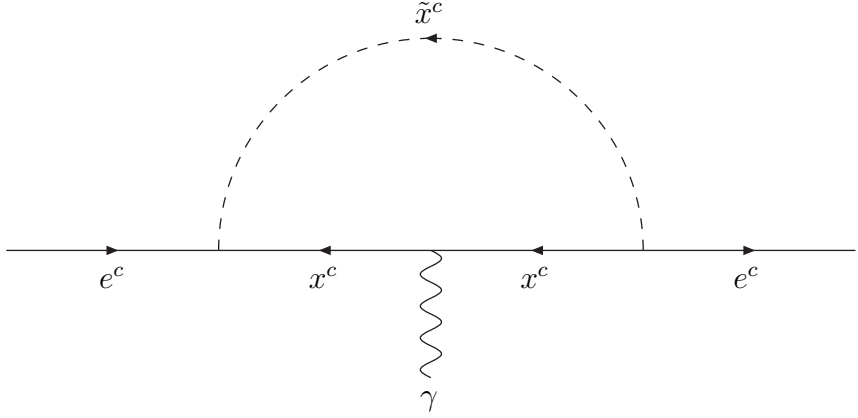

Using the Lagrangian of Eq.~(18), we obtain the amplitude for $l_i \to l_j 
+ \gamma$:
\begin{equation}
{\cal A} = \bar l_j (p-q) q^\mu \epsilon^\nu i \sigma_{\mu \nu} \left[ 
\xi \left( {1 + \gamma_5 \over 2} \right) + \eta \left( {1 - \gamma_5 \over 2} 
\right) \right] l_i(p),
\end{equation}
where
\begin{eqnarray}
\xi &=& {ie \over 64 \pi^2} \left[ {m_i \over M^2} f_1^i (f_1^j)^* [F(z_1) + 
z_1^{-1} F(z_1^{-1})] \right. \nonumber \\ && + \left. {m_i \over M_c^2} 
f_2^i (f_2^j)^* [F(z_2) + z_2^{-1} F(z_2^{-1})] + {m_j \over M_c^2} f_3^i 
(f_3^j)^* [F(z_3) + z_3^{-1} F(z_3^{-1})] \right],
\end{eqnarray}
and $\eta$ is obtained from $\xi$ with the interchange of $m_i$ and $m_j$, 
i.e. the masses of $l_i$ and $l_j$ respectively.  In the above, $M$ is the 
invariant fermion mass of the $x$ and $y$ hemions, and $M_c$ that of the 
$x^c$ and $y^c$ hemions.  The function $F(z)$ is given by
\begin{equation}
F(z) = {2 + 5z - z^2 \over 6 (1-z)^3} + {z \ln z \over (1-z)^4},
\end{equation}
where $z_1 = m^2_{\tilde y^c}/M^2$, $z_2 = m^2_{\tilde x}/M_c^2$, and 
$z_3 = m^2_{\tilde x^c}/M_c^2$.  In the limit $z=1$, $F(z)=1/12$.

As an illustration, let $z_1=z_2=z_3=1$, $f_1^{i,j}=f_2^{i,j}=f_3^{i,j}=f$, 
$M_c=M$, then the $\mu \to e \gamma$ rate is given by
\begin{equation}
\Gamma = {5 \alpha |f|^4 m_\mu^5 \over 9 (256)^2 \pi^4 M^4},
\end{equation}
yielding a branching fraction of
\begin{equation}
B = {5 \alpha \over 3072 \pi} \left( {|f|^2 \over G_F M^2} \right)^2.
\end{equation}
Using the present experimental upper bound \cite{mueg} of $1.2 \times 
10^{-11}$, we obtain
\begin{equation}
M > 7|f|~{\rm TeV}.
\end{equation}
For $|f|$ of order 0.1, $M$ may thus be of order 1 TeV, in keeping with the 
expectation of Eq.~(17).  It also shows that there is a good possibility 
for a future observation of $\mu \to e \gamma$ just below the present 
upper bound.

In conclusion, we have shown that the exotic fermions and scalars of 
half-integral charges, predicted in a natural extension of the Standard 
Model to include leptonic color and resulting in $SU(3)^4$ quartification, 
should have masses in the TeV range and may induce an observable $\mu \to 
e \gamma$ decay rate.

This work was supported in part by the U.~S.~Department of Energy
under Grant No.~DE-FG03-94ER40837.

\bibliographystyle{unsrt}

\end{document}